# The R-W Metric Has No Constant Curvature When Scalar Factor R(t) Changes With Time

## ---- The effluence on the Hubble constant, dark material and dark energy

**Mei Xiaochun**

**( Department of Physics, Fuzhou University, China )**

**Abstract** The real physics meaning of constant $\kappa$ in the Robertson-Walker metric is discussed when scalar factor R(t) is relative to time. Based on the curvature formula of the Riemannian geometry strictly, the spatial curvature of the R-W metric is $K = -(\dot{R}^2 + \kappa)/R^2$. It indicates that the spatial curvature of the R-W metric is not a constant when $\dot{R}(t) \neq 0$ and the constant $\kappa$ does not represent spatial curvature factor. It can only be considered as an adjustable parameter relative to the Hubble constant. For the expensive flat space, we would have $\kappa = -\dot{R}^2(t')$ at a certain moment $t'$. We would have $\kappa \neq 0$ in general situations. The result is completely different from the current understanding which is based on specious estimation actually, in stead of strict calculation. In light of this result, many conclusions in the current cosmology, such as the values of the Hubble constant, dark material and dark energy densities, should be re-estimated. In this way, we may get rid of the current puzzle situation of cosmology thoroughly.

## 1. When $\dot{R}(t) \neq 0$ the R-W metric has no constant curvature

Standard cosmology takes the Robertson-Walker metric as the basic frame of space-time. In light of the principle of cosmology, our universe is uniform and isotropic. It can be proved that the space with homogeneity and isotropy is one with constant curvature [1]. The R-W metric which is considered with a biggest symmetry is

$$ds^2 = c^2 dt^2 - R^2(t)\left( \frac{d\bar{r}^2}{1 - \kappa\bar{r}^2} + \bar{r}^2 d\theta^2 + \bar{r}^2 \sin^2\theta \, d\varphi^2 \right) \qquad (1)$$

In light of the current understanding, constant $\kappa$ in (1) represents spatial curvature factor. That is to say, the R-W metric has a constant curvature. When space is flat, we have $\kappa = 0$ and the metric becomes

$$ds^2 = c^2 dt^2 - R^2(t)\left( d\bar{r}^2 + \bar{r}^2 d\theta^2 + \bar{r}^2 \sin^2\theta \, d\varphi^2 \right) \qquad (2)$$

In light of the current understanding, because the part of three dimension space in (2) is lat, by multiplying a scalar factor $R(t)$ which having nothing to do with space coordinates, the space can still be considered flat. However, this is not true. We will prove strictly below that when $\dot{R}(t) \neq 0$, the R-W metric has no constant spatial curvature. The special curvature is relative to $\dot{R}(t)$, thought it still has nothing to do with spatial coordinates.

In order to see this clearly, let's first repeat the deducing process of the R-W metric. As we know in geology that the curved space of low dimension can be embedded into the flat space of high dimension. The three dimension space with a constant curvature $\kappa$ can be considered as a super-curved surface to embed into the four dimension flat space. The four dimension metric of flat space-time can be written as

$$ds^2 = dx_\mu dx_\mu = dx_i dx_i + (dx_4)^2 \qquad （3）$$

The super spherical surface condition of three dimensions in the four dimension space-time is

$$x_\mu x_\mu = x_1^2 + x_2^2 + x_3^2 + x_4^2 = x_1^2 + x_2^2 + x_3^2 - c^2 t^2 = \frac{1}{\kappa} = G^2 \qquad （4）$$

Here $G$ is the radius of super spherical surface and $\kappa =$ constant is the curvature of super spherical surface. By taking the differential of (4), we have

$$dx_4 = icdt = -\frac{x_i dx_i}{x_4} \qquad\qquad (dx_4)^2 = \frac{(x_i dx_i)^2}{1/\kappa - x_i x_i} = \frac{\kappa(x_i dx_i)^2}{1 - \kappa x_i x_i} \qquad （5）$$

By substituting (5) into (4) and introducing coordinate transformation $x_1 = r\sin\theta\cos\varphi$, $x_2 = r\sin\theta\sin\varphi$ and $x_3 = r\cos\theta$, we obtain the super-curved surface metric of three dimensions with a constant curvature $\kappa$ in the flat space-time of four dimensions

$$ds^2 = \frac{dr^2}{1 - \kappa r^2} + r^2 d\theta^2 + r^2 \sin^2\theta d\varphi^2 \qquad （6）$$

By introducing following coordinate $r = R(t)\bar{r}$, when time is fixed with $t = t_0$ and $R(t_0) =$ constant, (6) can be written as

$$ds^2 = R^2(t_0)\left( \frac{d\bar{r}^2}{1 - \kappa R^2(t_0)\bar{r}^2} + \bar{r}^2 d\theta^2 + \bar{r}^2 \sin^2\theta d\varphi^2 \right)$$

$$= R^2(t_0)\left( \frac{d\bar{r}^2}{1 - \kappa'\bar{r}^2} + \bar{r}^2 d\theta^2 + \bar{r}^2 \sin^2\theta d\varphi^2 \right) \qquad （7）$$

In which $\kappa' = \kappa R^2(t_0)$. In light of the current understanding, when time is not fixed with $R(t) \neq$ constant, let $\kappa' \rightarrow \kappa$, the formula (7) is extended into

$$ds^2 = R^2(t)\left( \frac{d\bar{r}^2}{1 - \kappa\bar{r}^2} + \bar{r}^2 d\theta^2 + \bar{r}^2 \sin^2\theta d\varphi^2 \right) \qquad （8）$$

For the expansive universe with homogeneity and isotropy, it is thought to be possible to introduce united time. So in the following coordinate system, the metric of four dimension space-time with a constant curvature for its three dimension space is written as

$$dS^2 = c^2 dt^2 - ds^2 = c^2 dt^2 - R^2(t)\left( \frac{d\bar{r}^2}{1 - \kappa\bar{r}^2} + \bar{r}^2 d\theta^2 + \bar{r}^2 \sin^2\theta d\varphi^2 \right) \qquad （9）$$

This is the standard process to reach the R-W metric.

It is obvious that the deduction of the W-R metric is not strict to contain some analogy and extending. What is verified strictly is only that the metric (6) has a constant curvature $\kappa$. When $R(t) \neq$ constant, we have not proved that the metrics (8) and (9) also has a constant curvature $\kappa$. In fact, when $R(t) \neq$ constant, substitute $r = R(t)\bar{r}$ into (6), we obtain

$$ds^2 = \frac{\dot{R}^2(t)dt^2}{1-\kappa R^2(t)\bar{r}^2} + \frac{2\dot{R}(t)R(t)dtd\bar{r}}{1-\kappa R^2(t)\bar{r}^2} + R^2(t)\left(\frac{d\bar{r}^2}{1-\kappa R^2(t)\bar{r}^2} + \bar{r}^2d\theta^2 + \bar{r}^2\sin^2\theta d\varphi^2\right) \quad (10)$$

The metric corresponding to (9) becomes

$$dS^2 = \left[c^2 - \frac{\dot{R}^2(t)}{1-\kappa R^2(t)\bar{r}^2}\right]dt^2 - \frac{2\dot{R}(t)R(t)dtd\bar{r}}{1-\kappa R^2(t)\bar{r}^2} - R^2(t)\left(\frac{d\bar{r}^2}{1-\kappa R^2(t)\bar{r}^2} + \bar{r}^2d\theta^2 + \bar{r}^2\sin^2\theta d\varphi^2\right) \quad (11)$$

We can not get (8) and (10) from (6). That is to say, when $\dot{R}(t) \neq 0$ or $R(t) \neq$ constant, we have not proved that the constant $\kappa$ is still the spatial curvature factor of the R-W metric! In fact, the metric of flat four dimension space-time is

$$ds^2 = c^2dt^2 - \left(dr^2 + r^2d\theta^2 + r^2\sin^2\theta \, d\varphi^2\right) \quad (12)$$

By introducing transformation $r(t) = R(t)\bar{r}$, we obtain

$$ds^2 = c^2\left[1 - \frac{\dot{R}^2(t)\bar{r}^2}{c^2}\right]dt^2 - 2R(t)\dot{R}(t)\bar{r}d\bar{r}dt - R^2(t)\left(d\bar{r}^2 + \bar{r}^2d\theta^2 + \bar{r}^2\sin^2\theta d\varphi^2\right) \quad (13)$$

Its form is completely different from the R-W metric (2) when $\kappa = 0$. In (13), space-time seems curved, but it is flat in essence. In light of the principle of the Riemannian geometry, if we can find a transformation to transform a curved space into flat, the original space is flat in essence. If we can not find such a transformation, the original space is curved really. Because we can find a transformation to change (13) into (12), so the metric (13) is flat actually. Because we can not find a transformation to change (2) into (12) when $\dot{R}(t) \neq 0$, the spatial part of (2) can not be flat!

The result above is completely different from the current understanding. Because it would cause great influence on cosmology, we should treat it seriously. We will calculate the spatial curvatures of the R-W metric based on the formula of the Riemannian geometry strictly and obtain a completely different result in next section. In view of the fact that the geometrical figure of supper spherical surface of three dimensions is indirect, in order to reach direct understanding, we discuss the spherical surface of two dimensions in three dimension's flat space further in this section. Similar to (3) and (4), the flat metric of three dimension's space and the condition of spherical surface are individually

$$d\sigma^2 = dx_1^2 + dx_2^2 + dx_3^2 \qquad x_1x_1 + x_2x_2 + x_3x_3 = \frac{1}{\kappa} = G^2 \quad (14)$$

Here constant $\kappa$ is the curvature and $G$ is the radius of spherical surface of two dimensions. By introducing column coordinates to let $x_1 = r\sin\theta$, $x_2 = r\cos\theta$ and $x_3 = z$, (14) becomes

$$d\sigma^2 = dz^2 + dr^2 + r^2d\theta^2 \qquad z^2 + r^2 = \frac{1}{\kappa} = G^2 \quad (15)$$

Here $r = \sqrt{x_1^2 + x_2^2}$ is the radius of a circle located on the plane in which $z$ is fixed. According to the same procedure, we have

$$dz = -\frac{rdr}{z} \qquad dz^2 = \frac{(rdr)^2}{1/\kappa - r^2} = \frac{\kappa(rdr)^2}{1-\kappa r^2} \quad (16)$$

$$d\sigma^2 = \frac{dr^2}{1-\kappa r^2} + r^2d\theta^2 \quad (17)$$

(17) is similar to (6). We see that (15) is flat, but (17) becomes curved. The reason is that transformation (16) is non-linear. That is to say, non-linear transformations would change spatial curvatures. Similarly, by introducing nonlinear transformation $r = R(z)\bar{r}$ and let $R'(z) = dR(z)/dz$ in (17), we obtain

$$d\sigma^2 = \frac{R'^2(z)\bar{r}^2 dz^2}{1 - \kappa R^2(z)\bar{r}^2} + \frac{R'(z)R(z)\bar{r}dzd\bar{r}}{1 - \kappa R^2(z)\bar{r}^2} + R^2(z)\left[\frac{d\bar{r}^2}{1 - \kappa R^2(z)\bar{r}^2} + \bar{r}^2 d\theta^2\right] \qquad (18)$$

The curvature of (18) can not be constant $\kappa$. Let

$$dS^2 = dz^2 + d\sigma^2 = dx_1^2 + dx_2^2 + 2dz^2 \qquad (19)$$

and substitute (18) in to (19), we obtain

$$dS^2 = \left[1 + \frac{R'^2(z)\bar{r}^2}{1 - \kappa R^2(z)\bar{r}^2}\right]dz^2 + \frac{R'(z)R(z)\bar{r}dzd\bar{r}}{1 - \kappa R^2(z)\bar{r}^2} + R^2(z)\left[\frac{d\bar{r}^2}{1 - \kappa R^2(z)\bar{r}^2} + \bar{r}^2 d\theta^2\right] \qquad (20)$$

(20) is similar to (11). The equations corresponding to (8) and (9) are

$$d\sigma^2 = R^2(z)\left(\frac{d\bar{r}^2}{1 - \kappa\bar{r}^2} + \bar{r}^2 d\theta^2\right) \qquad (21)$$

$$dS^2 = dz^2 + R^2(z)\left(\frac{d\bar{r}^2}{1 - \kappa\bar{r}^2} + \bar{r}^2 d\theta^2\right) \qquad (22)$$

It is obvious that (20), (21) and (22) can not have constant curvature $\kappa$ too. In fact, by introducing transformation $r = R(z)\bar{r}$, the equation of spherical surface (15) becomes

$$R^2(z)\bar{r}^2 + z^2 = \frac{1}{\kappa} \qquad (23)$$

Because new invariables are $\bar{r}$ and $z$, the formula is not the equation of spherical surface again. It becomes the equation of irregular curved surface without constant curvature. When $z = z_0 = $ constant, in light of (23), we have

$$\bar{r}^2 = \frac{1}{R^2(z_0)}\left(\frac{1}{\kappa} - z_0^2\right) = \frac{1}{K} \qquad (24)$$

Only under this condition, constant $K$ represent the curvature of a circle with radius $\bar{r}$ in a plane. Similarly, by considering $r = R(t)\bar{r}$, the equation (4) of supper spherical surface becomes

$$R^2(t)\bar{r}^2 - c^2 t^2 = \frac{1}{\kappa} \qquad (25)$$

Because (25) is not the equation of supper spherical surface of three dimension for new variables $\bar{r}$ and $t$, it has no constant curvature $\kappa$ too. Only when $t = t_0 = $ constant, we have

$$\bar{r}^2 = \frac{1}{R^2(t_0)}\left(\frac{1}{\kappa} - c^2 t_0^2\right) = \frac{1}{K} \qquad (26)$$

Constant $K$ represents the curvature of supper spherical surface with a radius $\bar{r}$.

## 2 The spatial curvature of the R-W metric

Now let's calculate the curvatures of (21), (22) and the R-W metric (9) strictly under condition $R(t) \neq$ constant. As we known that curvature has strict definition in mathematics. We should judge flatness of space by strict calculation, not only by apparent estimation. In the Riemannian geometry, the Riemannian curvature at a certain point of $N$ dimensions is defined as[2]:

$$K = \frac{R_{\alpha\beta\sigma\rho} p^\alpha q^\beta p^\sigma p^\rho}{(g_{\alpha\sigma} g_{\beta\rho} - g_{\sigma\beta} g_{\alpha\rho}) p^\alpha q^\beta p^\sigma p^\rho} \qquad (27)$$

The Riemannian curvature $K$ is relative to the selection of direction vectors $p^\alpha$ and $q^\beta$ at each point of space. We define covariant tensor $R_{\alpha\beta\sigma\rho} = R_{\alpha\beta\sigma}{}^\lambda g_{\lambda\rho}$ in which $R_{\alpha\beta\sigma}{}^\rho$ and $R_{\alpha\beta\sigma\rho}$ can be calculate by following formulas

$$R_{\alpha\beta\sigma}{}^\rho = \frac{\partial}{\partial x^\beta} \Gamma^\rho_{\sigma\alpha} - \frac{\partial}{\partial x^\alpha} \Gamma^\rho_{\sigma\beta} + \Gamma^\rho_{\lambda\beta} \Gamma^\lambda_{\sigma\alpha} - \Gamma^\rho_{\lambda\alpha} \Gamma^\lambda_{\sigma\beta} \qquad (28)$$

$$R_{\alpha\beta\sigma\rho} = \frac{1}{2} \left( \frac{\partial^2 g_{\rho\alpha}}{\partial x^\sigma \partial^\beta} + \frac{\partial^2 g_{\sigma\beta}}{\partial x^\rho \partial^\alpha} - \frac{\partial^2 g_{\alpha\sigma}}{\partial x^\rho \partial^\beta} - \frac{\partial^2 g_{\rho\beta}}{\partial x^\sigma \partial^\alpha} \right) + \Gamma_{\rho\alpha,\lambda} \Gamma^\lambda_{\sigma\beta} - \Gamma_{\rho\beta,\lambda} \Gamma^\lambda_{\sigma\alpha} \qquad (29)$$

$$\Gamma_{\alpha\beta,\sigma} = \Gamma^\lambda_{\alpha\beta} g_{\lambda\sigma} = \frac{1}{2} \left( \frac{\partial g_{\beta\sigma}}{\partial x^\alpha} + \frac{\partial g_{\sigma\alpha}}{\partial x^\beta} - \frac{\partial g_{\alpha\beta}}{\partial x^\rho} - \frac{\partial^2 g_{\rho\beta}}{\partial x^\sigma} \right) \qquad (30)$$

It is proved that if the curvatures are the same at all spatial points, i.e., $K = $ constant, we would have

$$K = \frac{R_{\alpha\beta\sigma\rho}}{g_{\alpha\sigma} g_{\beta\rho} - g_{\sigma\beta} g_{\alpha\rho}} \qquad (31)$$

Further, if space is flat, the Riemannian-Christoffel tensors would becomes zero everywhere with $R_{\alpha\beta\sigma}{}^\rho = 0$ or $R_{\alpha\beta\sigma\rho} = 0$.

Let's first calculate the curvature tensor of the metric (21). The non-zero metric tensors of (21) are

$$g_{11} = \frac{R^2}{1 - \kappa \bar{r}^2} \qquad g_{22} = R^2 \bar{r}^2 \qquad g^{11} = \frac{1 - \kappa \bar{r}^2}{R^2} \qquad g^{22} = \frac{1}{R^2 \bar{r}^2} \qquad (32)$$

The non-zero Christoffel signs are

$$\Gamma^1_{11} = -\frac{\kappa \bar{r}}{1 - \kappa \bar{r}^2} \qquad \Gamma^1_{22} = -(1 - \kappa \bar{r}^2) \bar{r} \qquad \Gamma^2_{12} = \frac{1}{\bar{r}} \qquad (33)$$

In light of (29), the only curvature tensor is

$$R_{1212} = -\frac{\kappa R^2 \bar{r}^2}{1 - \kappa \bar{r}^2} \qquad (34)$$

So according to the definition of (27), the Riemannian curvature of (21) is

$$K_{12} = \frac{R_{1212} p^1 q^2 p^1 q^2}{(g_{11} g_{22} - g_{12} g_{12}) p^1 q^2 p^1 q^2} = \frac{R_{1212}}{g_{11} g_{22}} = -\frac{\kappa}{R^2} \qquad (35)$$

In fact, let $\kappa \to \kappa R^2(z)$ in (21), we get (17). Then we calculate the curvature of (22). We consider $z$ as $x^0$, the metric tensors of (22) are

$$g_{00} = 1 \qquad g_{11} = \frac{R^2}{1 - \kappa \bar{r}^2} \qquad g_{22} = R^2 \bar{r}^2$$

$$g^{00} = 1 \qquad g^{11} = \frac{1 - \kappa \bar{r}^2}{R^2} \qquad g^{22} = \frac{1}{R^2 \bar{r}^2} \qquad (36)$$

The non-zero Christoffel signs are

$$\Gamma_{11}^0 = -\frac{RR'}{1-\kappa\bar{r}^2} \qquad\qquad \Gamma_{22}^0 = -RR'\bar{r}^2 \qquad\qquad \Gamma_{01}^1 = \Gamma_{02}^2 = \frac{R'}{R}$$

$$\Gamma_{11}^1 = -\frac{k\bar{r}}{1-\kappa\bar{r}^2} \qquad\qquad \Gamma_{22}^1 = (1-\kappa\bar{r}^2)\bar{r} \qquad\qquad \Gamma_{12}^2 = \frac{1}{\bar{r}} \qquad (37)$$

The non-zero curvature tensors are

$$R_{0101} = -\frac{RR''}{1-\kappa\bar{r}^2} \qquad\qquad R_{0202} = -RR''\bar{r}^2 \qquad\qquad R_{1212} = -\frac{R^2\bar{r}^2(R'^2+\kappa)}{1-\kappa\bar{r}^2} \qquad (38)$$

In light of (27), we have

$$K_{01} = \frac{R_{0101}}{g_{00}g_{11}} = -\frac{R''}{R} \qquad K_{02} = \frac{R_{0202}}{g_{00}g_{22}} = -\frac{R''}{R} \qquad K_{12} = \frac{R_{1212}}{g_{11}g_{22}} = -\frac{R'^2+\kappa}{R^2} \qquad (39)$$

$K_{12}$ in (39) is different from that in (35). It means that the spatial curvatures of (21) and (22) are different. The dimensions of space would change curvature, thought the metric form of two dimension curved surface in three dimension space shown in (22) is completely the same as (21). In fact, as we known that the metric $ds^2 = dr^2 + r^2 d\theta^2 + r^2\sin^2\theta\,d\varphi^2$ is flat with zero curvature, but $d\sigma^2 = r^2 d\theta^2 + r^2\sin^2\theta\,d\varphi^2$ is curved with curvature $\kappa = 1/r^2$, which is not a constant relative to the radius of sphere.

Now we discuss the space-time curvatures of the R-W metric (9), the non-zero curvature tensors can be calculated as below

$$R_{0101} = \frac{R\ddot{R}}{1-\kappa\bar{r}^2} \qquad\qquad R_{0202} = R\ddot{R}\bar{r}^2 \qquad\qquad R_{0303} = R\ddot{R}\bar{r}^2\sin^2\theta$$

$$R_{1212} = -\frac{R^2\bar{r}^2(\dot{R}^2+\kappa)}{1-\kappa\bar{r}^2} \qquad\qquad R_{1313} = -\frac{R^2\bar{r}^2(\dot{R}^2+\kappa)\sin^2\theta}{1-\kappa\bar{r}^2}$$

$$R_{2323} = -R^2\bar{r}^4(\dot{R}^2+\kappa)\sin^2\theta \qquad (40)$$

The corresponding curvatures are

$$K_{01} = K_{02} = K_{03} = -\frac{\ddot{R}}{R} \qquad\qquad K_{12} = K_{13} = K_{23} = -\frac{\dot{R}^2+\kappa}{R^2} \qquad (41)$$

Here $K_{0i}$ are the curvature of space-time cross part of the R-W metric, and $K_{ik}$ ($i,k\neq 0$) are the curvatures of pure space. In order to reach $K_{0i}=0$, we should have $\ddot{R}=0$ or $\dot{R}=$ constant. In this case, $K_{ik}$ are not constant. To let spatial curvature metric $K_{ik}=0$, we have both chooses. One is to let $\dot{R}=0$ and $\kappa=0$ simultaneously. In this case, we also have $K_{0i}=0$. Another is to let $\kappa=-R^2(t')$ at a certain moment $t=t'$. In this case, we have $K_{ik}=0$ but still have $K_{0i}\neq 0$. Only in theses two cases, the spatial part of the R-W metric can be considered flat. If only take $\kappa=0$, we have $K_{ik}=-\dot{R}^2/R^2\neq 0$, the spatial part of the R-W metric can not be flat. The result is completely different from the current understanding which is based on specious estimation, in stead of strict calculation. Only when $\dot{R}(t)=0$ or $R=$ constant, we obtain the result of current theory with $K_{ik}=-\kappa/R^2$.

Now let's estimate the magnitude of the curvatures of the R-W metric. By taking $\kappa=0$ in (41), we obtain

$$K_{12} = K_{13} = K_{23} = -\frac{\dot{R}^2}{R^2} \tag{42}$$

We define $\dot{R}(t)/R(t) = H(t)$. At present moment $t_0$ we have the Hubble constant $H_0 \sim 2 \times 10^{-18} s^{-1}$. So we have $K_{1i} \sim -4 \times 10^{-36}$. From the equation of cosmology, we have $\ddot{R}(t_0)/R(t_0) \sim 4\pi G\rho(t_0)/3 \sim H_0^2/2 = 2 \times 10^{-36} s^{-1}$, so we have $K_{0i} \sim -2 \times 10^{-36}$. Therefore, in light of the R-W metric with $\kappa = 0$, even though the space of the expansive universe is curved, the curvature is also very small with a magnitude $10^{-36}$.

Then, how can we recognize the spatial curvature of (2)? In the Riemannian geometry, the intuitionistic picture is that when a vector moves along a loop and returns to original point, if the moving vector can superpose with original vector, the space would be considered flat. If it can not, the space would be curved. So, it means that when a vector moves along a loop and returns to the original point $(\bar{r}, \theta, \varphi)$, it can not superpose with original vector in light of (2) during a period of time.

The result of WMAP showed that the space of our universe seems flat[3]. So in light of (41), the difference between constant $\kappa$ and the value of $\dot{R}^2(t_0)$ at present moment can not be too great. If we think that constant $\kappa$ with the same magnitude of $\dot{R}^2(t_0)$, the spatial curvature of the R-W metric would be about $10^{-36}$. The precision of WMAP is about $10^{-3}$, so the experiment can not find such small spatial curvature. On the other hand, the observations of WMAP are carried out through measuring the isotropy of cosmic microwave background radiations, which were found having nothing to do with angles $\theta$ and $\varphi$. But if space curvature is only relative to time, having nothing to do with spatial coordinate $r$ $\theta$ and $\varphi$, WMAP may be unable to find the curvature shown in (41).

On the other hand, in order to make the spatial part of the R-W metric keeps positive, or to make the distance to be a positive number, we should have

$$\frac{R^2(z)d\bar{r}^2}{1 - \kappa\bar{r}^2} \geq 0 \tag{43}$$

So we must have $1 - \kappa\bar{r}^2 > 0$, i.e., we must have $\kappa < 0$ or $0 < \kappa < 1/\bar{r}^2$. Because $\bar{r}$ may be a very big number, for example, the radius of the observable universe with $\bar{r} \sim 10^{26}$ m, we would have $\kappa < 10^{-52}$. That is to say, $\kappa$ should be a very small positive number. If the universe is infinite, we would have $\kappa \to 0$. In the current cosmology, we take $\kappa = -1$，0 and 1, to represent the universe with negative curvature, flatness and positive curvature. However, $\kappa = 1$ is impossible. If taking $\kappa = 1$, we can only take $\bar{r} < 1$ to ensure (43) meaningful. In the region $\bar{r} > 1$, the R-W metric becomes meaningless in practical measurement. Unfortunately, this problem is completely neglected in current cosmology.

## 3. Constant $\kappa$ in the Friedmann equation

In order to see the real meaning of constant $\kappa$ in the R-W metric further, we discuss the strict solution of the Einstein's equation of gravity with spherical symmetry and material uniform distribution. The solution is the Schwarzschild metric which is divided into two sections, inner solution and external solution. For a static and uniform sphere with radius $r_1$, suppose that material density $\rho_0$ and intensity of pressure $p_0$ are constants. By considering static energy momentum tensor of ideal liquid, the Schwarzschild metric of inner sphere is[4]:

$$ds^2 = c^2 \left( \frac{3}{2}\sqrt{1 - \frac{8\pi G\rho_0 r_1^2}{3c^2}} - \frac{1}{2}\sqrt{1 - \frac{8\pi G\rho_0 r^2}{3c^2}} \right)^2 dt^2 - \left( \frac{dr^2}{1 - 8\pi G\rho_0 r^2/3c^2} + r^2 d\theta^2 + r^2 \sin^2\theta d\varphi^2 \right) \quad (44)$$

If we take $r_1$ as the radius of the visible universe with $r_1 \sim 10^{26}$, (44) can be considered to describe the static and uniform universe approximately. If take $\rho_0 = 0$, (44) becomes the metric of flat space-time. By introducing the Hubble constant $H_0 = \sqrt{8\pi G\rho_0/3}$ and let $\kappa = H_0^2/c^2$, we can write (44) as

$$ds^2 = c^2 \left( \frac{3}{2}\sqrt{1 - \kappa r_1^2} - \frac{1}{2}\sqrt{1 - \kappa r^2} \right)^2 dt^2 - \left( \frac{dr^2}{1 - \kappa r^2} + r^2 d\theta^2 + r^2 \sin^2\theta d\varphi^2 \right) \quad (45)$$

Let $r_1 \to r$, (45) becomes (1). By using transformation $r = R(t)\bar{r}$, (45) becomes

$$ds^2 = c^2 \left\{ \left( \frac{3}{2}\sqrt{1 - \kappa'\bar{r}_1^2} - \frac{1}{2}\sqrt{1 - \kappa'\bar{r}^2} \right)^2 - \frac{\dot{R}^2(t)\bar{r}^2}{c^2(1 - \kappa'\bar{r}^2)} \right\} dt^2$$

$$- \frac{2R(t)\dot{R}(t)\bar{r}dtd\bar{r}}{c(1 - \kappa'\bar{r}^2)} - R^2(t) \left( \frac{d\bar{r}^2}{1 - \kappa'\bar{r}^2} + \bar{r}^2 d\theta^2 + \bar{r}^2 \sin^2\theta d\varphi^2 \right) \quad (46)$$

Here $\kappa'(t) \sim H^2(t)R^2(t)/c^2 \neq$ constant. So we can consider $\kappa \to \kappa'(t_0)/R^2(t_0) = H_0^2/c^2$ as the real physical significance of constant $\kappa$ in the R-W metric.

The different understanding of constant $\kappa$ in the R-W metric would cause great influence in cosmology. By considering cosmic constant $\lambda$, the Friedmann equation of cosmology is written as

$$\frac{\dot{R}^2}{R^2} + \frac{\kappa}{R^2} + \lambda = \frac{8\pi G}{3}\rho \quad (47)$$

If constant $\kappa \neq 0$, it would cause great influence on the values on the Hubble constant, dark material and dark energy. If taking $\kappa = -\dot{R}^2(t_0)$ at present time, we would have $\lambda = 8\pi G\rho(t_0)/3$. In light of (41), the space is flat in this case. The cosmic constant is equal to material density. By defining the Hubble constant $H = \dot{R}(t)/R(t)$, we can write the Friedmann equation as

$$\rho_e(t) = \frac{3\dot{R}^2(t)}{8\pi GR^2(t)} + \frac{3\kappa}{8\pi GR^2(t)} = \frac{3H^2(t)}{8\pi G} + \frac{3\kappa}{8\pi GR^2(t)} \quad (48)$$

Here $\rho_e(t) = \rho(t) - 3\lambda/(8\pi G)$ is called as the effective density of the universal material. Let $\rho_e(t_0) = \rho_0$, $H(t_0) = H_0$, $R(t_0) = R_0 = 1$ and define critical density $\rho_c = 3H_0^2/(8\pi G)$, (48) becomes

$$\rho_0 = \rho_c + \frac{3\kappa}{8\pi G} \quad (49)$$

The formula is used to estimate the material density of the universe. If $\kappa$ represents space curvature factor, we have $\kappa = 0$ for the flat universe so that current material density is equal to critical density. By defining $\mathbf{\Omega} = \rho/\rho_c$, we would have $\mathbf{\Omega}_0 = \rho_0/\rho_c = 1$ for the current universe. However, observations show that we only have $\Omega_0 = 0.04$ for normal material, which is greatly less than 1, so that non-baryon dark material and dark energy are needed to fill the universe. If constant $\kappa \neq 0$, we can chose proper $\kappa$ to satisfy (47) and (49). At present, the Hubble constant is considered to be $H_0 = 65 Km \cdot s^{-1} \cdot Mpc^{-1}$, so we have $\rho_c = 7.9 \times 10^{-27} kg/m^3$. By the practical measurements, we have $\rho_0 \approx 2 \times 10^{-28} kg/m^3$ for laminate material[5]. Suppose that practical density of baryon material in the universe is about 10 times more than laminate material, we can take $\kappa = 3.3 \times 10^{-36}$ to satisfy (47). In this way, it becomes unnecessary for us to suppose that the non-baryon material is about 6 times more than baryon material in the universe if they exist really. As for the hypothesis of dark energy, it would becomes surplus. Theory and observations can be consistent by adjusting the value of constant $\kappa$ without the hypotheses of dark energy. On the other hand, if $\kappa$ does not represent spatial curvature, in light of the Friedmann equation, we have

$$H^2(t) = \frac{8\pi G \rho(t)}{3} - \frac{\kappa}{R^2(t)} \qquad (50)$$

In this case, the Hubble constant is not only relative to material density, but also relative to scalar factor. As mentioned before, we should have $\kappa < 0$. For the current moment, we have $R_0 = 1$, so we obtain

$$H_0 = \sqrt{\frac{8\pi G \rho_0}{3} - \frac{\kappa}{R_0^2}} = \sqrt{\frac{8\pi G \rho_0}{3} - \kappa} > \sqrt{\frac{8\pi G \rho_0}{3}} \qquad (51)$$

In fact, as we know that the Hubble constant is a quite complex quantity, not a pure constant actually. It is obvious that many problems in cosmology are relative to the real meaning of constant $\kappa$. If $\kappa$ is not a curvature factor and can be considered as an adjustable parameter, many conclusions in the current cosmology should be re-estimated.

# 4. Discussion

It is proved that constant $\kappa$ in the R-W metric can not represent space curvature factor when scalar factor R(t) is relative to time. The real spatial curvature of the R-W metric is $K = -(\dot{R}^2 + \kappa)/R^2$ based on the strict calculation of the Riemannian geometry. Therefore, if the Friedmann equation is used to describe the expensive universe, we can only consider $\kappa$ as an adjustable and non-zero parameter which is relative to the Hubble constant. The result would cause great influence on cosmology. We need to re-estimate the values of the Hubble constant, dark material and dark energy densities. According to the current estimation, for the universal material, normal baryon material only makes up $4\%$, non-baryon dark material makes up $26\%$ and dark energy makes up $70\%$. However, no non-baryon dark material with so much quantity can be founded up to now though physicists have struggled for decades. The situation is more puzzling for dark energy.

In fact, the concept of dark energy is similar to the concept of ether in the nineteen century. Based on the Newtonian theory, classical physics could not explain the light's propagation in the universal vacuum. The medium of ether had to be introduced. One of purposes that Einstein put forward special relativity was to drive out ether from physics. Time goes into the twenty-one century, the history is repeating. Physicists can not explain the high red-shift of supernovae, the concept of dark energy has to be introduced. By the correct understanding of constant $\kappa$ in the Friedmann equation, we do not need the concept of dark energy again. Meanwhile, we may not need to suppose that non-baryon dark material is 6 times more than normal baryon material, if non-baryon dark material exists actually. In this way, we may get rid of the current puzzle situation of cosmology thoroughly.